\documentclass[aps,prl,superscriptaddress,twocolumn,showpacs]{revtex4}

\usepackage{amsmath}
\usepackage{graphicx}
\begin{document}
\title{Instability and Nonlinear Evolution of Narrow-Band Directional Ocean Waves}
\author{Bengt Eliasson and P. K. Shukla}
\affiliation{Fakult\"at f\"ur Physik und Astronomie,
Ruhr--Universit\"at Bochum, D-44780 Bochum, Germany, and Department of Physics,
Ume\aa~ University, SE-90187 Ume\aa,~ Sweden}
\received{2 December 2009}
\revised{3 June 2010}

\begin{abstract}
The instability and nonlinear evolution of directional ocean waves is investigated numerically
by means of simulations of the governing kinetic equation for narrow-band
surface waves. Our simulation results reveal the onset of the modulational instability for long-crested
wave-trains, which agrees well with recent large-scale experiments in
wave-basins, where it was found
that narrower directional spectra leads to self-focusing of ocean waves and an enhanced probability of extreme events. We find that the modulational instability is nonlinearly saturated by a broadening of the wave-spectrum,
which leads to the stabilization of the water-wave system. Applications of our results to other fields
of physics, such as nonlinear optics and plasma physics are discussed.
\end{abstract}

\maketitle

Giant freak waves, or rogue waves, have been observed in mid-ocean and coastal
waters \cite{Kharif03}, in optical systems \cite{Solli07}, and in parametrically driven
capillary waves \cite{Shats10}. The freak/rogue waves are short-lived phenomena appearing
suddenly out of normal waves, and with a small probability \cite{Akhmediev09}.
The study of extreme gravity waves on the open ocean has important applications for the sea-faring
and offshore oil industries, where they may lead to structural damage and injuries
to personnel \cite{Kharif03}.
It is, therefore, very important to understand the physical mechanisms that lead to
the formation of freak waves. Since the linear theory cannot explain the number of
extreme events that occur in the ocean and in optical systems, one has to account
for nonlinear effects (e.g. wave-wave interactions) in combination with the wave dispersion.
This can lead to the modulational instability (for water waves called the
Benjamin-Feir instability  \cite{Benjamin67,Ruban07}), followed by focusing and amplification
of the wave energy.

Wind-driven waves on the ocean often have wide frequency spectra that are peaked
in the direction of the wind \cite{JONSWAP,Mitsuyasu75,Hasselmann80}. The statistics of directional
spectra for narrow-band gravity waves have also recently been studied experimentally in water
basins \cite{Onorato09,Onorato09b,Waseda09}, where it was found that sea states with narrow directional
spectra (long--crested waves) were more likely to produce extreme waves. Examples of statistical models
that govern collective interactions of groups of water waves are Hasselmann's model \cite{Hasselmann62}
for random, homogeneously distributed waves and Alber's model \cite{Alber78} for narrow-banded wave trains.
Wave-kinetic simulations in one spatial dimension have shown Landau damping and coherent
structures \cite{Onorato03}, and recurrence phenomena \cite{Stiassnie08} for random water wave fields.
In this Letter, we derive a nonlinear wave-kinetic (NLWK) equation for gravity waves in $2+2$
dimensions (two spatial dimensions and two velocity dimensions) and carry out simulations to study
the stability and nonlinear spatio-temporal evolution of narrow-band spectra waves that were
observed in the recent experiments by Onorato and coworkers \cite{Onorato09}.  The present NLWK model,
which is similar to Alber's model \cite{Alber78}, is particularly suitable for studying the
nonlinear dynamics of narrow-band water waves due to its relative simplicity. Similar nonlinear
wave-kinetic equations also appear in the description of optical systems, photonic lattices,
and plasmas \cite{Bingham97}.

Deep water gravity waves are governed by the dispersion relation $\omega=\sqrt{g k}$,
where $g$ is the gravitational constant, $k=\sqrt{k_x^2+k_y^2}$ is the modulus of the wave vector
${\bf k}=k_x \widehat{\bf x}+k_y\widehat{\bf y}$, and $\widehat{\bf x}$ and $\widehat{\bf y}$ are
the unit vectors in the $x-$ and $y-$directions. Assuming surface displacements of the form $\eta=(1/2)A({\bf r},t)\exp(-i\omega_0 t+i k_0 x)$+ complex conjugate, where $A$ is the slowly varying ($|\partial/\partial t|\ll\omega_0$, $|\nabla|\ll k_0$) envelope, ${\bf r}=x \widehat{\bf x}+y\widehat{\bf y}$ is the spatial coordinate, and $\omega_0=\sqrt{g k_0}$, the nonlinear interaction of water waves is governed by the nonlinear Schr\"odinger equation (NLSE)
%1
\begin{equation}
   i\left( \frac{\partial A}{\partial t}+v_{gr}\frac{\partial A}{\partial x}
   \right)+D_x\frac{\partial^2 A}{\partial x^2}+D_y\frac{\partial^2 A}{\partial y^2}-\xi|A|^2 A=0,
  \label{NLSE}
\end{equation}
where $v_{gr}=\partial \omega/\partial k_x=\omega_0/2 k_0$ is the group velocity,
$D_x=(1/2)\partial^2\omega/\partial k_x^2=-\omega_0/8k_0^2$ and
$D_y=(1/2)\partial^2\omega/\partial k_y^2=\omega_0/4k_0^2$ are the group dispersion coefficients,
and the nonlinear coupling coefficient is $\xi=\omega k_0^2/2$.
%The NLSE is valid
%for waves with wave-vectors close to $k_0\widehat{\bf x}$ and the corresponding frequencies
%close to $\omega_0=\sqrt{g k_0}$.
Introducing the two-dimensional Wigner transform \cite{Moyal}
%2
\begin{equation}
  f({\bf r}, {\bf v}, t)=\frac{1}{2(2\pi)^2} \int A^*({\bf R}_{+},t)A({\bf R}_{-},t)
e^{i{\boldsymbol\lambda}\cdot({\bf v}-v_{gr}\widehat{\bf x})}\,d^2\lambda,
\label{Wigner_transform}
\end{equation}
where we have denoted ${\bf R}_\pm={\bf r}\pm{\bar{\bar{\bf D}}}\cdot{\boldsymbol{\lambda}}$
and ${\bar{\bar{\bf D}}}\cdot{\boldsymbol{\lambda}}=D_x\lambda_x\widehat{\bf x}+D_y\lambda_y\widehat{\bf y}$,
we obtain the evolution equation for the pseudo-distribution function $f$ as
%3
\begin{equation}
  \begin{split}
  \frac{\partial f}{\partial t}&+{\bf v}\cdot\nabla f-\frac{2 i \xi}{(2\pi)^2}\int\int
    [I({\bf R}_{+},t)-I({\bf R}_{-},t)]
    \\
  &\times~f({\bf r},{\bf v}',t)e^{i{\boldsymbol\lambda}\cdot({\bf v}-{\bf v'})}d^2v'\,d^2\lambda=0,
  \end{split}
  \label{Wigner}
\end{equation}
where $I({\bf r},t)=\int f({\bf r},{\bf v},t)\,d^2v$ is the variance of the surface displacement (the wave intensity).
The transformation (\ref{Wigner_transform}) between (\ref{NLSE}) and (\ref{Wigner}) is valid in both
directions for a deterministic wave-train (corresponding to a ``pure state'' in quantum mechanics),
with some restrictions on the distribution function $f$ \cite{Moyal}; however, we are interested in
the statistical properties of an ensemble of waves, and more general choices of $f$ where the
deterministic picture is abandoned \cite{Alber78}. In the absence of the nonlinear term in the left-hand
side of (\ref{Wigner}), we have ${\partial f}/{\partial t}+{\bf v}\cdot\nabla f=0$, which dictates that
the wave energy propagates in space with the group velocity ${\bf v}$. Our model is valid for waves
with ${\bf v}\approx v_{gr}\widehat{\bf x}$. The dispersive properties of the wave are
important for the nonlinear wave-wave interactions between wave-packets that are modeled by
the interaction integral in the last term in the left-hand side of (\ref{Wigner}).

The velocity distribution can be related to the wave spectrum in frequency domain.
Similar to Ref. \cite{Onorato09}, we will use the model spectrum parameterized by the
Joint North Sea Wave Project (JONSWAP) as \cite{JONSWAP}
%4
\begin{equation}
  S(\omega)=\frac{\alpha_P g^2}{\omega^{5}}\exp\left(-\frac{5}{4}\frac{\omega_p^4}{\omega^4}\right)
  \gamma^{\exp\left[-\frac{(\omega-\omega_p)^2}{2\sigma^2\omega_p^2}\right]},
  \label{PM_spectrum}
\end{equation}
where $\omega_p$ is the peak frequency, $\gamma$ is the peak enhancement parameter and $\alpha_P$
is the Phillips parameter. Here $\gamma$ is in the range 1--6 for ocean waves \cite{Onorato09},
while $\alpha_P$ is in the range $0.0081$--$0.1$; the values $\gamma=1$ and $\alpha_P=0.0081$
gives the spectrum of fully developed wind seas \cite{Pierson64}, while the larger values
are observed in water tank experiments. We will use $\alpha_P \approx 0.025$, $\gamma=3$
and $\sigma=0.08$, which are consistent with the Marintek water basin experiment
in Refs.~\cite{Onorato09,Onorato09b}.  Since the wave spectrum is concentrated around $\omega=\omega_p$,
we will use $\omega_0=\omega_p$ and $k_0=k_p\equiv\omega_p^2/g$ in the evaluation of $D_x$ and $D_y$
in Eq.~\ref{Wigner}.

The integral of the spectrum (\ref{PM_spectrum}) over all frequencies yields the variance of the surface elevation.
While the variance of a monochromatic wave is $|A|^2/2$, from (\ref{Wigner_transform}) we also have
$\int f\,d^2v=|A|^2/2$. Hence, as initial conditions in our simulations, we will use $f=f_0({\bf v})=F_0(v)G(\theta)$  where we have introduced polar coordinates  $v_x=v\cos(\theta)$ and $v_y=v\sin(\theta)$ in velocity
space. We obtain $F_0$ from the frequency spectrum (\ref{PM_spectrum}) by using the differential
variance $dI=S(\omega)d\omega= F_0({v})\,vdv$, as
%5
\begin{equation}
  F_0(v)=S[\omega(v)]\frac{1}{v}\left|\frac{d\omega}{d v}\right|=S[\omega(v)]\frac{g}{2 v^3},
\end{equation}
where we used that the group speed $v$ of the wave packets
is related to the wave frequency $\omega=\sqrt{gk}$ via $v={d \omega}/{dk}=\omega/2k=g/2\omega$, or
$\omega(v)={g}/{2v}$. The directional spreading function is chosen as \cite{Mitsuyasu75}
$G(\theta)=G_0 \cos^{N}({\theta}/{2})=G_0 {[1+\cos(\theta)]^{N/2}}/{2^{N/2}}$,
where $\cos(\theta)=v_x/v$, $v=(v_x^2+v_y^2)^{1/2}$,
and $G_0$ is a normalization constant \cite{Mitsuyasu75} such that $\int_{-\pi}^{\pi}G(\theta)\,d\theta=1$.
We note that $G$ has a maximum at $\theta=0$ and tends to a narrower distribution with an increase
of the parameter $N$.

\begin{figure}[htb]
\includegraphics[width=8.5cm]{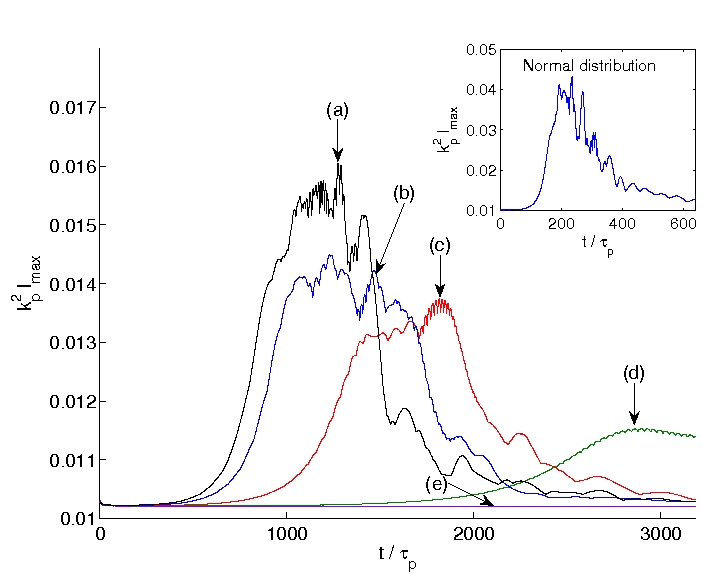}
\caption{(Color online) The time-evolution of the maximum intensity $k_p^2 I_{max}$ for
(a) $N=840$ (black)
(b) $N=200$ (blue),
(c) $N=90$ (red),
(d) $N=50$ (green),
(e) $N=24$ (magenta), and the case of a narrow-band normal velocity distribution (the inset).
 The spatial distributions of wave intensity for (a)--(d)  are shown
in Fig.~\ref{fig3} at the times indicated here with arrows.}
\label{fig1}
\end{figure}

\begin{figure}[htb]
\includegraphics[width=8.5cm]{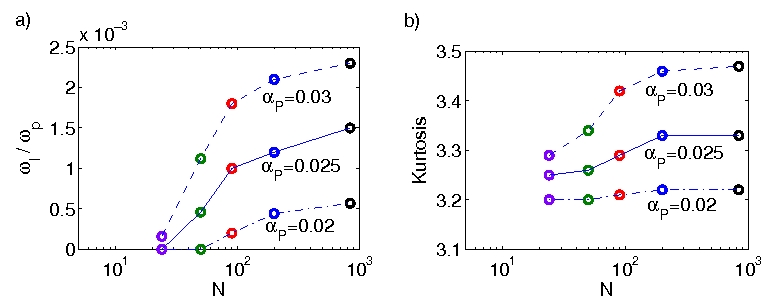}
\caption{(Color online) a) The linear growth rate $\omega_I$ of the fastest growing wavemode
and b) maximum kurtosis for $N=840$ (black) $N=200$ (blue), $N=90$ (red), $N=50$ (green), and $N=24$ (magenta),
for $\alpha_P=0.02$ (dashed line), $\alpha_P=0.025$ (solid line) and $\alpha_P=0.03$ (dash-dotted line).
The solid line ($\alpha_P=0.025$) corresponds to curves (a)--(e) in Fig.~\ref{fig1}.}
\label{fig2}
\end{figure}

Equation (\ref{Wigner}) can be cast into a numerically more convenient form
by employing the Fourier-transform in velocity space
%6
\begin{equation}
  \widehat{f}({\bf r},{\boldsymbol\eta},t)=2\int f({\bf r},{\bf v},t)e^{i{\boldsymbol\eta}\cdot{\bf v}}\,d^2v,
  \label{Fourier}
\end{equation}
which transforms Eq. (\ref{Wigner}) into
%7
\begin{equation}
  \frac{\partial \widehat{f}}{\partial t}-i\nabla_{\boldsymbol\eta}\cdot\nabla \widehat{f}
  +2i\xi[I({\bf r}+{\bar{\bar{\bf D}}}\cdot{\boldsymbol{\eta}},t)
-I({\bf r}-{\bar{\bar{\bf D}}}\cdot{\boldsymbol{\eta}},t)]\widehat{f}({\bf r},{\boldsymbol\eta},t)=0,
  \label{Wigner2}
\end{equation}
where $I=\widehat{f}({\bf r},{\boldsymbol\eta},t)_{{\boldsymbol\eta}={\boldsymbol0}}/2$.
A similar equation was derived by Alber \cite{Alber78}, starting from the Davey-Stewartson equations for weakly nonlinear gravity waves.
The numerical approximation of (\ref{Wigner2}) is based on a method to solve the Fourier transformed Vlasov
equation \cite{Eliasson02}. Using a pseudo-spectral method in space, the operator $\nabla$ is converted
to multiplication by $i{\boldsymbol \kappa}$, and the spatial shifts by $\pm{\bar{\bar{\bf D}}}\cdot{\boldsymbol\eta}$
in Eq.~(\ref{Wigner2}) are converted to multiplications by
$\exp[\pm i ({\bar{\bar{\bf D}}}\cdot{\boldsymbol\eta})\cdot {\boldsymbol \kappa}]$,
where ${\boldsymbol \kappa}$ is the wave vector.  The system was solved in a computational
window moving with the group speed of the peak wave.
We used a spatial domain of size $L_x\times L_y=100\,k_p^{-1}\times 500\,k_p^{-1}$, resolved
by $N_x\times N_y=32\times 32$ intervals and with periodic boundary conditions, and a Fourier transformed
velocity domain $L_{\eta x} \times L_{\eta y}= 160\pi v_{ph}^{-1}\times 160\pi v_{ph}^{-1}$
with $N_{\eta x}\times N_{\eta y}=80 \times 80$ intervals, where $v_{ph}=\omega_p/k_p$ is the phase speed
of the peak wave. The velocity domain in our simulations is thus $v_{x,min}\leq v_x\leq v_{x,max}$ and
$v_{y,min}\leq v_x\leq v_{y,max}$ where $v_{x,min}=0$, $v_{x,max}= 2\pi N_{\eta x}/L_{\eta x}=2 v_{gr}$,
and $-v_{y,min}=v_{y,max}=\pi N_{\eta y}/L_{\eta y}=v_{gr}$. The simulation was initialized with the JONSWAP
spectrum, where the Fourier integral (\ref{Fourier}) was evaluated numerically to obtain the spectrum
in $\boldsymbol{\eta}$ space. Random numbers of the order $10^{-2}$ of the initial intensity was added
to the solution in order to seed the modulational instability. The initial conditions give an intensity of
$I\approx 0.010\, k_p^{-2}$ uniformly distributed in space, which is compatible with the experiments of
Onorato {\it et al.} \cite{Onorato09}. To compare with the experimental observations of
Onorato {\it et al.} \cite{Onorato09}, we carried out simulations for $N=24$, $50$, $90$, $200$,
and $840$ corresponding to the Marintek experiment in Ref.~\cite{Onorato09}. They
used $\omega_p=2\pi\,\mathrm{s}^{-1}$ (1 Hz) and corresponding $k_p= 4.1\,\mathrm{m}^{-1}$,
and a significant wave height $H_s=0.08\,\mathrm{m}$, giving a wave intensity of
$I\approx 5\times10^{-4}\,\mathrm{m}^2$.

\begin{figure}[htb]
\includegraphics[width=8.5cm]{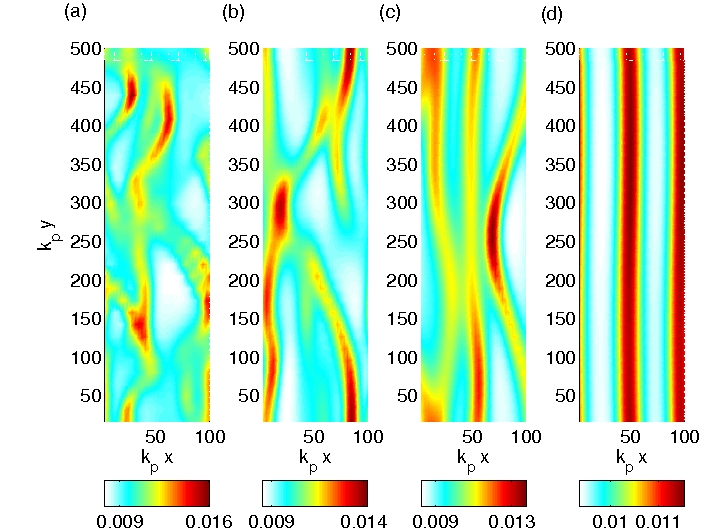}
\caption{(Color online) The spatial distribution of the normalized wave intensity $k_p^2 I$ for
(a) $N=840$ at $t=1.27\times10^3\,\tau_p$,
(b) $N=200$ at $t=1.46\times10^3\,\tau_p$,
(c) $N=90$ at $t=1.81\times10^3\,\tau_p$, and
(d) $N=50$ at $t=2.90\times10^3\,\tau_p$, corresponding
to the curves (a)--(d) in Fig.~\ref{fig1}.}
\label{fig3}
\end{figure}

\begin{figure}[htb]
\includegraphics[width=8.5cm]{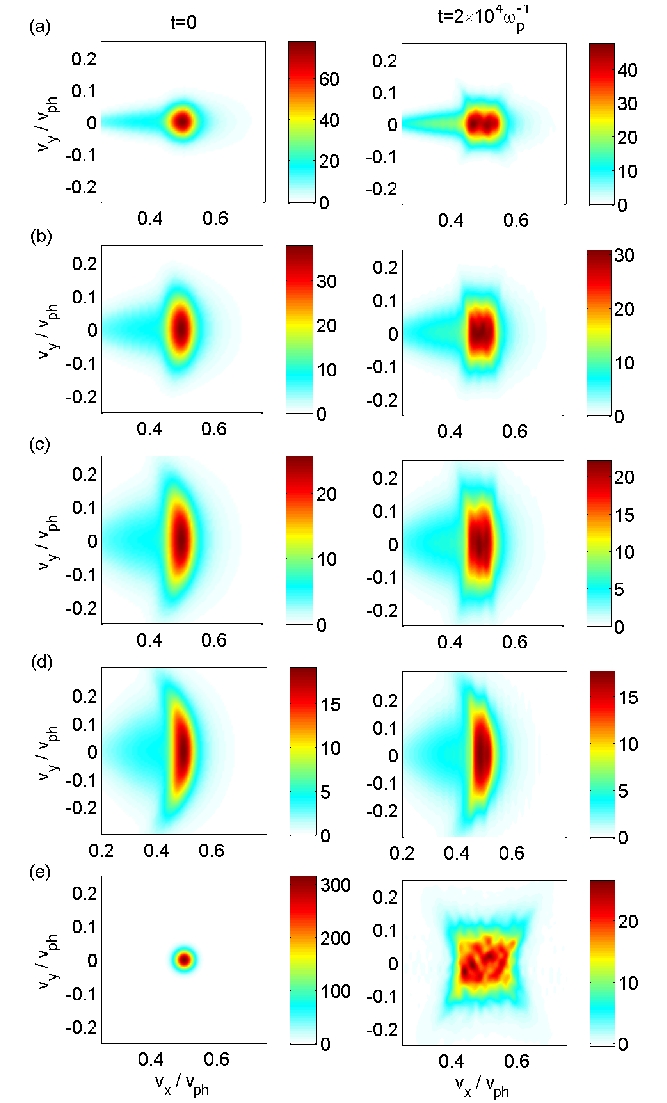}
\caption{(Color online) The velocity distribution $\omega_p^2 f$ of the wave energy, averaged over space,
at $t=0$ (left column) and $t=3.2\times10^3\tau_p$ (right column), for (a) $N=840$
(b) $N=200$, (c) $N=90$, (d) $N=50$. Panel (e) shows the narrow-band normal velocity
distribution at $t=0$ (left) and $t=640\,\tau_p$ (right).}
\label{fig4}
\end{figure}

According to the analysis of Alber \cite{Alber78}, using a model two-dimensional normal spectrum,
there are two conditions for the modulational instability: first, the modulational wavenumbers must lie
within a certain directional range (in Alber's case $|K_x|>\sqrt{2}|K_y|$ similar to the Benjamin-Feir instability),
and second, the wave steepness (the wave amplitude multiplied by $k_p$) must be larger than
the normalized (by the component of the spectral peak) spectral bandwidth. In our simulations, using directional JONSWAP spectra, we observed
the modulational instability and the self-focusing of the wave energy into localized wave packets for
$N$ larger than $24$. We measured the maximum value of the energy density in the simulation domain,
and plotted its time evolution in Fig.~\ref{fig1} (the time is give in units of the peak wave
period $\tau_p=2\pi/\omega_p$). Initially, there is an exponential growth phase, reminiscent of the Benjamin-Feir
instability for monochromatic wave trains \cite{Benjamin67}. The modulational instability is fastest growing
for $N=840$, and decreases with decreasing values of $N$. For $N=24$ we do not observe any
instability.  For modulationally unstable cases, the exponential growth phase is followed by a nonlinear saturation
of the instability, and finally a decrease of the maximum energy density down to its initial background
value $I\sim 0.01 k_p^{-2}$, as seen in curves (a)--(d) of Fig.~\ref{fig1}. The inset shows a simulation with
a narrow-band normal distribution of the form $f=4\omega_p^{-2}\exp[-2 (v_y^2+(v_x-v_gr)^2)/\sigma^2]$
with $\sigma=0.04\,v_{ph}$, which yields the initial wave intensity $I=0.01 k_p^{-2}$ that is similar
as in curves (a)--(d). This case shows a rapidly growing instability to large amplitudes and then a decrease.
The linear growth rate $\omega_I$ of the instability for different values of $N$ and $\alpha_P$ was measured
from the data and plotted in Fig. \ref{fig2}(a).  The growth rate is larger up to some limiting value for long-crested
waves with $N>10^2$, while it approaches zero for smaller values of $N$.
A growth rate of $\omega_I=1$--2$\times10^{-3}\omega_p$ implies
an amplitude doubling of the unstable wave in 50--100 wave periods. The growth rate is sensitive to changes
of $\alpha_P$ and shows an increase/decrease of 50\% with an increase/decrease of $\alpha_P$ by 20\%; this is
consistent with a ratio of unity between the wave steepness and the spectral bandwidth, so that the system is weakly unstable. The strongly unstable case for the narrow normal distribution has
a growth rate $\omega_I\approx 0.008\omega_p$, which is close to the limiting value \cite{Alber78}
$\omega_I=I k_p^2 \omega_p$ for monochromatic waves.

The kurtosis is traditionally \cite{Higgins63} estimated by the formula $\lambda_4=3+24 k_p^2 \sigma^2$,
where $\sigma$ is the standard deviation of the surface elevation. (The factor 3 comes from the assumption of Gaussian statistics and the term $24 k_p^2 \sigma^2$ is a nonlinear correction to the Gaussian statistics.) Assuming that the wave field is ergodic,
we have $\sigma^2=\langle I\rangle$, where $\langle I\rangle$ is the spatially averaged wave intensity.
As noted in Ref. \cite{Onorato09b}, this formula underestimates the kurtosis compared to the experimental
values for narrow-band water waves, where an increase of the kurtosis was observed at later stages of the wave dynamics.
Our model also conserves $\langle I\rangle$ and hence the formula predicts constant kurtosis.
Taking into account that the wave-field is non-stationary and that the wave intensity varies in space (see Fig. 3), we, instead, estimate the kurtosis as $\lambda_4=3\langle I^2 \rangle/\langle I \rangle^2+24 k_p^2 \langle I \rangle$, which assumes that the surface obeys Gaussian statistics locally everywhere. Using this estimate, we see in Fig. 2(b) that larger $N$ gives larger kurtosis, in good agreement with experimental observations \cite{Onorato09,Onorato09b,Waseda09}. Figure \ref{fig3} shows that the wave energy is concentrated
into narrow bands, elongated along the $y$-direction, which are propagating from left to
right with speeds close to $v_{gr}$.
At later stages, the wave-packets start to break up due to the two-dimensionality in space and
the elongated bands of wave energy become more and more wiggled with the appearance of obliquely propagating waves, similar to those observed in Ref. \cite{Ruban07}. For the modulationally unstable cases, the
nonlinear interaction leads to a broadening of the distribution function in velocity space, as seen in Fig. \ref{fig4}. This, in turn, leads to a stabilization of the system via phase mixing
of the wave envelopes \cite{Alber78}, and a saturation and decrease of the maximum intensity shown in Fig.~1.

To summarize, we have performed a series of kinetic simulations of narrow-banded water
waves for different degrees of directional energy spectra. We observe an onset of the modulational instability and self-focusing of the wave energy for waves with narrow directional spectra,
leading to an increase of the estimated kurtosis.
The modulational instability saturates via the occurrence of narrow wave-packets, which later disperse due
to the broadening of the wave spectrum.  Our simulation results are in excellent agreement with observations
from recent large-scale experiments in wave-basins \cite{Onorato09,Onorato09b,Waseda09}.

\end{document}